\documentclass[12pt]{article}

\usepackage{arxiv}
\usepackage{hyperref}       
\usepackage{url}            
\usepackage{booktabs}       
\usepackage{amsmath}        
\usepackage{amsfonts}       
\usepackage{microtype}      
\usepackage{graphicx}
\usepackage{mathptmx}
\usepackage[frozencache,cachedir=minted-cache]{minted}         
\usepackage{tikz}
\usepackage{indentfirst}
\usetikzlibrary{shapes,arrows}
\usetikzlibrary{calc,decorations.pathmorphing}

\tikzstyle{block} = [draw, fill=blue!20, rectangle, minimum height=3em, minimum width=6em, text width=6em, align=center]
\tikzstyle{dot} = [circle,fill,blue!80,inner sep=0pt,minimum size=1.5mm]

\tikzset{
  radiation/.style={{decorate,decoration={expanding waves,angle=90,segment length=4pt}}},
  eye/.pic={
    \def\topedge{(-3,0) .. controls (-2,1.8) and (2,2) .. (2.3,.3)}
    \def\bottomedge{(2.3,.3) .. controls (2,-2.2) and (-2,-1.2) .. (-3,0)}
    \def\eyepath{\topedge -- \bottomedge --cycle;}
    \clip\eyepath;
    \filldraw[color=orange!30!black] (-.2,.2) circle (1.4);
    \filldraw[color=orange!40!black] (-.3,-.1) circle (1.1);
    \foreach \a in {0,5,...,360}{
      \pgfmathparse{25+28*rnd}
      \fill[orange!\pgfmathresult!black, decoration={random steps, segment length=1pt, amplitude=0.3pt}, decorate, line width=0.3pt  ] (-.2,.2) -- ++($(\a+2*rnd:.8+0.3*rnd)$) -- ++(\a+90:3pt) -- cycle;
    }
    \fill[color=black] (-.2,.2) circle (0.5);
    \draw[line width=2.5mm, draw opacity=0.1, line cap=round]\topedge;
    \draw[line width=1mm, red!40!white!80!black, line cap=round]\bottomedge;
    \draw[line width=1.2mm, red!40!white!60!black, line cap=round]\topedge;
    \fill[red!40!white!80!black] (-2.8,0) circle (.25);
    \fill[white] (-2.7,.1) circle (.03);
}}

\usepackage[backend=biber,hyperref=true,style=ieee,bibencoding=utf8]{biblatex}
\makeatletter
\def\blfootnote{\xdef\@thefnmark{}\@footnotetext}
\makeatother

\title{Automated Logging Drone: A Computer Vision Drone Implementation}
\author{Aaron Yagnik\\ Cornell University\\ \url{ay294@cornell.edu}}
\author{
    Aaron Yagnik\thanks{This work was done while the first author was an intern at Synechron Inc., supervised by the second author.}\\
    Cornell University\\
    \texttt{ay294@cornell.edu} \\
    \And
    \href{https://orcid.org/0000-0002-9067-5663}{\includegraphics[scale=0.06]{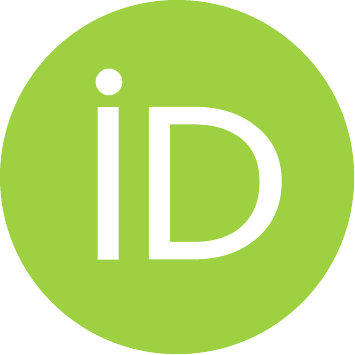}\hspace{1mm}Adrian S.-W. Tam}\\
    Synechron Inc.\\
    \texttt{adrian.tam@synechron.com}
}
\date{} 
\renewcommand{\undertitle}{}   
\renewcommand{\headeright}{}   

\begin{document}

\maketitle

\section*{Abstract}
In recent years, Artificial Intelligence (AI) and Computer Vision (CV) have become the pinnacle of technology with new developments seemingly every day. This technology along with more powerful drone technology have made autonomous surveillance more sought after. Here an overview of the Automated Logging Drone (ALD) project is presented along with examples of how this project can be used with more refining and added features.

\section{Introduction}

Drone technology nowadays is matured. Relatively inexpensive drones are available off-the-shelf and already come with sophisticated and reliable control. Therefore it is no surprise to see drones used increasingly in the world in a multitude of fields from food and package delivery to military surveillance all over the world.

Computer vision technology advanced significantly in the past decade. Not only did the computational performance open the door to more practical use, but also various new techniques improved the accuracy of computer vision to make it more reliable. We also saw that various computer vision models are available on the internet in the form of libraries or source code, which allows the model to be integrated to a larger system easily.

The motivation of this project is a simple one: In this post-Covid time, the office occupancy is not as high as previous years and is likely to remain as such. It would be useful to have a device rather than a human to patrol around the office. Even better if it is not remotely controlled but automatic. Using drones together with computer vision is a solution to this. Should there be a meeting or a visit to the office, we can let the drone verify the occupancy of the office. While a security camera can give you perspective from a fixed location, a drone allows us to explore a wider area and take a closer look at point of interest on demand.

Using a drone with preset flight path can partially fulfill the goal of remote site surveillance, however obstacles may appear in the preset path (e.g., a helium balloon left behind from a party the day before). Moreover, the preset flight may subject to turbulence which is difficult to correct without visual feedback. Thus the study of how we can use computer vision with a drone would be pertinent. We named this project the \emph{automated logging drone} (ALD) as the drone will be connected to a computer over wireless connection and to which the image captured by the drone camera will be delivered and optionally recorded on the computer.

\section{Set up}
\begin{figure}
    \centering
    \includegraphics{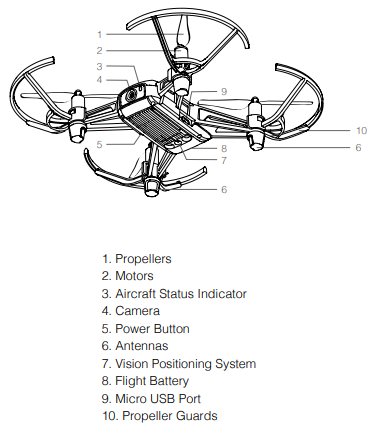}
    \caption{Aircraft diagram \cite{tello}}
    \label{fig:aircraft}
\end{figure}

In order for this to be possible without building a machine from scratch, the drone had to be open-source and controllable with a programming library. After searching through hundreds of different models, we chose to use the DJI Tello/Tello EDU drones (\figurename~\ref{fig:aircraft}). DJI is a company that develops and builds drones with state-of-the-art technology. The Tello drones can be controlled by a Python library named \emph{djitellopy}. Since djitellopy is a Python library, We are also able to use OpenCV, MediaPipe, TensorFlow, NumPy, and so on. This drone seemed to be the best option in terms of mission compatibility.

The Tello drone has some simplistic features that both help and restrict the capabilities of the ALD. It has an onboard forward-facing camera with capabilities of capturing JPG photos and MP4 videos. It also carries a Visual Positioning System (VPS) comprised of a monochromatic downward-facing camera and a 3D infrared module on the bottom of the body. It also carries a strong flight controller and multiple Wi-Fi antennas that allows safe control up to distances of 100 meters. The 100 meter range should be sufficient in indoor environment since we may extend the Wi-Fi reception range with extenders or mesh networks. However, this drone is a small one with limited lifting power. It is not designed to carry any load. Moreover, it needs good ventilation to operate as it shuts down automatically when it overheats at around 95\textdegree{}F. Thus the drone should not explore hot areas such as spaces near heat dissipating machinery.

The drone has a 2.4 GHz 802.11n Wi-Fi access point built-in. This is the means by which a computer connects to it. The drone and the computer communicate via UDP over Wi-Fi. This includes the drone receiving control signal from the computer as well as delivering the image captured from the camera to the computer.

While the drone is equipped with VPS, we found that the djitellopy library cannot access the VPS. This restricted the capabilities of the drone as we cannot make it follow a flight path marked on the ground. However, the front-facing camera works and this is sufficient to implement the ALD.

\section{Design}

At a high level, we are implementing a feedback control to the drone. With the djitellopy library to connect to the drone, we can control the movement of the drone by assigning the \emph{speed} to four axes, namely, the $x$, $y$, and $z$ axes in a 3D Cartesian coordinate system as well as the rotation about the vertical ($z$) axis. We named these left-right, forward-backward, up-down, and yaw.

The drone is controlled in terms of speed because it is directly related to the power delivered to the motors. However, this means we cannot precisely control the displacement directly. In fact, the drone's movement is not always precise and the error is perceivable, especially in the vertical direction. Therefore, we would combine the visual information from the camera to create a feedback control loop to achieve desired displacement.

\begin{figure}[h]
    \centering
    \begin{tikzpicture}[auto, node distance=3cm, >=latex', font=\footnotesize]
        \coordinate (orig) at (0,0);
        \node [block, left of=orig] (drone) {Drone};
        \node [block, right of=orig] (algo) {Control Algorithm};
        \coordinate [above of=orig, node distance=15mm] (dronedata);
        \coordinate [below of=orig, node distance=15mm] (controldata);
        \draw [->] (drone) |- (dronedata) node[above] {camera image} -| (algo);
        \draw [->] (algo) |- (controldata) node[below] {speed adjustments: $x$, $y$, $z$, yaw} -| (drone);
    \end{tikzpicture}

    \caption{Feedback control of ALD}\label{fig:feedback}
\end{figure}
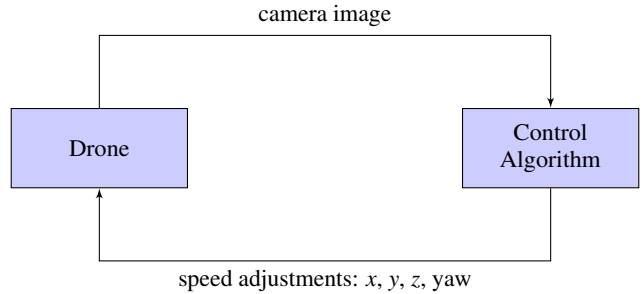

The high level feedback control is illustrated in \figurename~\ref{fig:feedback}. Making the drone follow a human face, for example, would need the drone to provide the image as it sees in the camera and from it, we can derive how the drone should move toward goal of maintaining a constant distance from the human. However, adjusting the speed of the drone does not affect the image from the camera immediately because of the reaction time in the drone as well as the round trip time of delivering the image. Therefore the feedback loop imposed a small delay in each iteration to avoid over reacting.

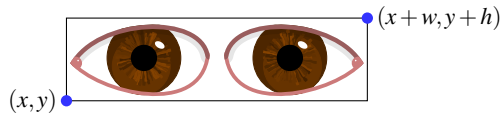
\begin{figure}[h]
    \centering
    \begin{tikzpicture}[auto, node distance=3cm, >=latex', font=\footnotesize]
        \coordinate (LL) at (0,0.5);
        \coordinate (UR) at (4,1.6);
        \draw (LL) node[dot] {} -| (UR) node[dot] {} -|cycle;
        \node [anchor=east] at (LL) {$(x,y)$};
        \node [anchor=west] at (UR) {$(x+w,y+h)$};
        \draw (1.1,1) pic[scale=0.35] {eye};
        \begin{scope}[shift={(1.1,1)},scale=0.35]
            \fill[color=white] (90:.8) {[rotate=-30] circle (0.2 and 0.12)};
        \end{scope}
        \draw (2.9,1) pic[xscale=-0.35,yscale=0.35] {eye};
        \begin{scope}[shift={(3.0,1)},scale=0.35]
            \fill[color=white] (90:.8) {[rotate=-30] circle (0.2 and 0.12)};
        \end{scope}
    \end{tikzpicture}

    \caption{A bounding box identified by the coordinates of the oppsite corners}\label{fig:bbox}
\end{figure}

The principal goal of this project is to make the drone follow a human. To reduce the complexity of implementation, we decided to assume there is only one human face visible from the camera and we look for and detect the eyes. For every frame we captured from the camera, we locate the eyes in a rectangular bounding box. A bounding box is identified by the coordinate of the two opposite corners (\figurename~\ref{fig:bbox}). With the bounding box $\{(x,y), (x+w,y+h)\}$, we can infer the box area $A=wh$ and the box center $C=(x+\frac{w}{2}, y+\frac{h}{2})$. Then we can compare $A$ to a threshold to determine if the drone should go forward or backward, since closer (the eyes and the bounding box) is bigger. Similarly, the $y$-coordinate of box center (relative to the image center) helps determine the vertical movement. We do not move the drone horizontally, but instead, we adjust the yaw (rotation about the vertical axis) based on the $x$-coordinate of the box center. Note that, from each image captured by the camera, we determined these three movements simultaneously and feedback to the drone.

The design of not moving the drone horizontally but to rotate it allows the drone to better follow the target (eyes of a human face) and adjust its orientation. Regarding the bounding box center $C$, We noticed that horizontal movement and yaw both adjust $C$ along the same direction. Therefore, using only one makes our feedback system work unambiguously.

\section{Discussion}

In this section, we discuss the trade-off made and our experience in developing the ALD system.

\paragraph{Development Hints} The drone allows us to control its speed at the motors but how the power output at motors translates into the movement of the drone depends on many factors, including the ambient air flow and the drone's momentous speed. In our program, we added a mechanism to use keyboard to control the motor speeds directly (mimicking a joystick control) so we can learn how the adjustment to the speeds reflect on the displacement of the drone. This mechanism also gives us a safety trigger to manually shut off the drone immediately if required.

For a similar reason, we implemented a mechanism to use hand gesture to control drone movements as well as applying real-time face recognition to drone camera so we learn about the performance of implementing computer vision to the control loop, referenced to \cite{hand,face,ytdji}. In this regard, we display a window to show the video as captured by the drone's camera and draw the control data, such as bounding box, as the on-screen display. We found this helpful to debug the control loop design.

\paragraph{Image Resolution} The drone returns the video as captured by the camera in the form of images of each frame, which each image is stored as a NumPy array. It is trivial to save a video file from the frames using OpenCV. However, we notice that computer vision features can be computationally expensive and significantly impact the control performance.

We used MediaPipe for hand gesture recognition. We found it fast enough to be part of our control system. However, some performance tweaks were still required. The drone camera gives us image captures in $840\times720$ pixel resolution. For identifying hand gestures presented in the image, a pixel resolution of $210\times 180$ (i.e., a quarter size in each dimension or $\frac{1}{16}$ in number of pixels) is sufficient. Therefore, our computer vision pipeline is as follows: For each image captured from the drone camera, we make a copy of it in the reduced dimension and submit to the computer vision library (e.g., MediaPipe) and get the result. The result is then transformed back to the original dimension for annotation (e.g., drawing the bounding box) on the original image. Shall we need multiple steps for computer vision (e.g., facial recognition and hand gesture recognition), the same reduced resolution image can be reused intact as the annotation is drew on the original image. Using the images in their original resolution would be slow and introduce a significant delay in the feedback control loop. This would be an impediment to the responsiveness and accuracy of the system.

\paragraph{Computer Vision Techniques} MediaPipe is backed by TensorFlow. It provides good quality result but the memory footprint is quite large and the computation burden is moderately heavy. We used MediaPipe for hand gesture recognition and it can also be used for facial recognition. Nevertheless, it is not the only tool available. In our implementation, we used Haar cascades \cite{haar,haar2} from OpenCV \cite{opencv} for facial recognition and eyes detection instead of Convolutional Neural Networks as used in MediaPipe. The Haar cascade is computationally less expensive and the difference in real-time performance is noticeable.

We implemented the drone to follow a human and initially we implemented this using facial recognition. However, it appears that facial recognition sometimes mistook certain background objects as faces which disrupted the drone's movement, and occasionally misled the drone to move to an unsafe direction. Therefore, we revised it to track eyes instead of the entire face as we saw it produced less false positive recognition in practice. In fact, in over 20 trials, the drone never once recognized another object as a pair of eyes.

\paragraph{Future Work} In our program we adjust the speed of the drone in a fixed step. The API from djitellopy allows the speed to be specified in the range of integers from $-100$ to $100$ but when we track the eyes, we emit only $-30$ or $30$ to move the drone forward or backward. This makes our implementation easy, since the control decision is binary, but undoubtedly allowing the speed to set faster or slower can make the drone move to its desired position in a shorter time. This would require a more sophisticated algorithm such as a Kalman filter to determine the feedback, however.

A further enhancement to the system is to let the drone identify other objects, such as a specific QR code or a picture. A more useful enhancement is to make the drone less sensitive to lighting conditions, which we noticed affects the recognition accuracy significantly.

\section{Conclusion}

We implemented a drone system with computer vision in the feedback loop to deliver automatic movement. The feedback mechanism is valuable to make the drone movement precise.

We identified a few techniques essential to create a successful drone system. This small drone system can find its use in many scenarios. One example is to take inventory of warehouses and making predictions of when supply will run out. This is similar to our motivation of patrol around the office. Another example is mapping areas affected by natural disasters which, if equipped with object recognition (such as identifying power lines, pipes, and people), can provide useful information to direct rescue crews and allocate relief resources. We can also make this system survey real estate to help construction industries or to assist in real estate transactions.

\section*{Acknowledgements}
The author would like to acknowledge the support of Mr. Sandeep Kumar of Synechron Inc. for his support and inspiration of this project without which this would have never been possible.

\appendix

\section{Complete code to the drone control system}

\begin{minted}[fontsize=\footnotesize]{python}
import logging
import threading
import time

import cv2
import numpy as np
import pygame
import tensorflow as tf
from djitellopy import tello
from mediapipe.solutions import hands as mphands
from mediapipe.solutions import drawing_utils as mputils

# Logging set up - global for this program
logging.basicConfig()
logger = logging.getLogger("drone")
logger.setLevel("DEBUG")

# computer vision and machine learning models
eye_cascade = cv2.CascadeClassifier("haarcascade_mcs_eyepair_big.xml")
face_cascade = cv2.CascadeClassifier("haarcascade_frontalface_default.xml")
gesture_names = [ln.strip() for ln in open("gesture.names").readlines()]
hand_keypoints = mphands.Hands(max_num_hands=1, min_detection_confidence=0.75)
gesture_classifier = tf.keras.models.load_model('mp_hand_gesture')

# global parameters
RES_DISP = (840, 720)   # camera image display resolution in (width,height)
RES_QUART = (RES_DISP[0]//4, RES_DISP[1]//4)   # 1/4 resolution for object detection
TRACK_WEIGHT = [0.4, 0.4]     # eye tracking coefficients


def getkey(keyname) -> bool:
    "check if the specified key is pressed down"
    for event in pygame.event.get():
        pass
    mykey = getattr(pygame, f'K_{keyname}')
    keyinput = pygame.key.get_pressed()
    is_pressed = bool(keyinput[mykey])
    pygame.display.update()
    return is_pressed


def record_video(controller):
    """record video from the camera on the drone to a MP4 file on disk,
    assumed to run in a separate thread"""
    fps = 30
    stream = controller.drone.get_frame_read()
    height, width, _ = stream.frame.shape
    with cv2.VideoWriter("video.mp4v", cv2.VideoWriter_fourcc(*'XVID'), fps, (width, height)) as vid:
        while controller.recording:
            vid.write(stream.frame)
            time.sleep(1.0/fps)


#
# Computer vision and image processing functions
#

def detect_eyes(image) -> tuple[int, int, int]:
    """detect the location of eyes from the image. The image will be annotated
    for the eye and the pixel coordinate of the eye will be returned. If there
    are multiple candidate, the one with the largest area will be used.

    Returns:
        area, center_x, center_y
    """
    # use opencv's Haar cascade to detect eyes
    img_gray = cv2.cvtColor(image, cv2.COLOR_BGR2GRAY)
    candidates = []
    for x, y, w, h in eye_cascade.detectMultiScale(img_gray, 1.2, 8):
        # red bounding box on each pair of eyes and mark center with green dot (BGR color system)
        cv2.rectangle(image, (x, y), (x+w, y+h), (0, 0, 255), 2)
        cx, cy = x + (w//2), y + (h//2)
        cv2.circle(image, (cx, cy), 5, (0, 255, 0), cv2.FILLED)
        # represent candidate as area and center coordinate
        candidates.append((w*h, cx, cy))
    # return the best candidate
    if not candidates:
        return 0, 0, 0
    return max(candidates)


def detect_face(img_big, img_small):
    """detect the location of faces from the image

    Args:
        img_big: the original image, annotation will be added
        img_small: the original image resized to 1/4 in dimension
    """
    # use opencv's Haar cascade to detect faces
    img_gray = cv2.cvtColor(img_small, cv2.COLOR_BGR2GRAY)
    faces = face_cascade.detectMultiScale(img_gray, 1.2, 8)
    num_faces = len(faces)
    # annotate each face with a bounding box
    for x, y, w, h in faces:
        # scale back to dimension of the original image
        x, y, w, h = 4*x, 4*y, 4*w, 4*h
        cv2.rectangle(img_big, (x, y), (x+w, y+h), (0, 0, 255), 2)

    # write the face count on image
    white = (255,255,255)
    coord1 = (120, 120)  # coord. to write "Faces detected:"
    coord2 = (150, 150)  # coord. to write the number
    cv2_put_text(img_big, "Faces detected:", coord1, white)
    cv2_put_text(img_big, str(num_faces), coord2, white)
    return img_big


def detect_gesture(img_big, img_small):
    """detect the hand gesture from image

    Args:
        img_big: the original image, annotation will be added
        img_small: the original image resized to 1/4 in dimension
    Returns:
        the best detected gesture
    """
    sm_w, sm_h, sm_c = img_small.shape
    # convert to RGB and get hand keypoint prediction
    rgb_small = cv2.cvtColor(img_small, cv2.COLOR_BGR2RGB)
    result = hand_keypoints.process(cv2.flip(rgb_small, 1))
    # annotate the keypoints detected: coordinates are based on a mirrored image
    img_big = cv2.flip(img_big, 1)   # flip for annotation
    gestures = []
    for hand in result.multi_hand_landmarks:
        # draw on frame
        mputils.draw_landmarks(img_big, hand, mphands.HAND_CONNECTIONS)
        # take the 21 keypoints in pixel coordinates
        landmarks = [[int(lm.x*sm_w), int(lm.y*sm_h)] for lm in hand.landmark]
        # infer gesture using keypoint coordinates
        proba = gesture_classifier([landmarks])
        classid = np.argmax(proba)
        gestures.append((proba[classid], classid))
    img_big = cv2.flip(img_big, 1)   # flip it back after annotation
    # return the best-confidence hand gesture
    _, classid = max(gestures)
    classname = gesture_names[classid]
    cv2_put_text(img_big, classname, (10, 50), (0,0,255))
    return classname


#
# On-screen display functions
#

def cv2_put_text(image, text, coord, color):
    """write text on image just like cv2.putText() but with black shadow"""
    font, flag, scale, thickness = cv2.FONT_HERSHEY_SIMPLEX, cv2.LINE_AA, 1, 2
    black = (0,0,0)
    text_x, text_y = coord
    shadow_coord = (text_x - 3, text_y - 3)
    cv2.putText(image, text, shadow_coord, font, scale, black, thickness, flag)
    cv2.putText(image, text, coord, font, scale, color, thickness, flag)
    return image


def put_temperature(image, temp):
    """Write the drone temperature on image, with color"""
    color = (255,255,255) if temp < 90 else (0,0,255)   # red if overheat, white otherwise
    coord1 = (400, 100)
    coord2 = (400, 150)
    cv2_put_text(image, "Internal Drone Temp:", coord1, color)
    cv2_put_text(image, str(temp), coord2, color)
    return image


def put_gesture_status(image, gesture_on):
    """Write the gesture control status"""
    coord = (370, 680)
    white = (255,255,255)
    msg = "Gesture Control Active" if gesture_on else "Gesture Control Inactive"
    cv2_put_text(image, msg, coord, white)
    return image


def put_follow_status(image, follow_on):
    """Write the follow status"""
    coord = (60, 680)
    white = (255,255,255)
    msg = "Follow Me Active" if follow_on else "Follow Me Inactive"
    cv2_put_text(image, msg, coord, white)
    return image


class DroneControl:
    """drone control system"""
    def __init__(self):
        "initialize state, I/O system, and connection to the drone"
        pygame.init()
        _ = pygame.display.set_mode((400, 400))
        self.drone = tello.Tello()
        self.drone.connect()
        logger.info(self.drone.get_battery())
        # state variables
        self.terminate = False     # signal to terminate this program
        self.recording = False     # video is recorded to file
        self.streaming = False     # drone camera should stream to a opencv window
        self.followme  = False     # drone should follow using eye-tracking
        self.gestures  = False     # gesture controlled drone
        self.findfaces = False     # count the number of faces on camera
        self.prev_offset = 0       # last x-offset for eye-tracking
        # launch the recording thread - always recording until controller terminates
        self.recorder = threading.Thread(target=record_video, args=(self,))

    def start_recording(self):
        "turn on the camera on the drone and stream video to a local window"
        self.drone.streamon()
        # get first image to start a opencv window, refresh_camera() will update it later
        img = self.drone.get_frame_read().frame
        img = cv2.resize(img, RES_DISP)
        cv2.imshow("Live Drone Video", img)
        cv2.waitKey(1)
        self.recorder.start()

    def end_recording(self):
        "turn off the camera on the drone and recall the recording thread"
        self.drone.streamoff()
        self.recorder.join()

    def refresh_camera(self):
        """get the latest frame from the drone camera and display on opencv
        window, as well as sending new control signal to the drone"""
        img = self.drone.get_frame_read().frame
        img_full = cv2.resize(img, RES_DISP)
        img_small = cv2.resize(img, RES_QUART)
        movement = None

        if self.findfaces:
            # find number of faces and mark on image
            detect_face(img_full, img_small)
        if self.gestures:
            # gesture control: get the gesture and move the drone
            gesture = detect_gesture(img_full, img_small)
            movement = self.fly_by_gesture(gesture)
        if self.followme:
            # find the location of eyes on image and move the drone
            area, cx, cy = detect_eyes(img_full)
            movement = self.fly_by_eyetracking(area, cx, cy)

        # write operation status on image
        put_temperature(img_full, self.drone.get_temperature())
        put_gesture_status(img_full, self.gestures)
        put_follow_status(img_full, self.followme)
        cv2.imshow("Live Drone Video", img_full)
        cv2.waitKey(1)
        return movement

    def fly_by_gesture(self, gesture) -> tuple[int, int, int, int]:
        """Translate gesture into flying instructions"""
        mapping = {     # RIGHT FORWARD UPWARD YAW
            "thumbs up":   [0,     0,     50,   0],   # move up
            "thumbs down": [0,     0,    -50,   0],   # move down
            "fist":        [0,    50,      0,   0],   # move forward
            "stop":        [0,   -50,      0,   0],   # move backward
        }
        if gesture in mapping:
            return mapping[gesture]
        else:
            return [0,0,0,0]

    def fly_by_eyetracking(self, area, x, y, area_range=(2800,3000)) -> tuple[int,int,int,int]:
        """
        Args:
            area: the area of the bounding box for the eye
            x, y: the pixel location of the center of the bounding box for the eye
            area_range: normal range for the bounding box area, this evaluates whether
                        the drone is too close or too far to the eyes
        """
        # yaw to left and right: compute horizontal offset of the eye w.r.t. center of the image
        w, h = RES_DISP
        if x > 0:
            # compare current eye offset to previous offset (in pixels) to determine movement on time axis
            x_offset = x - w / 2
            yaw = TRACK_WEIGHT[0] * x_offset + TRACK_WEIGHT[1] * (x_offset - self.prev_offset)
            yaw = int(np.clip(yaw, -100, 100) / 1.5)
        else:
            x_offset = yaw = 0
        # forward/backward: compare the size of the eyes to the normal range
        if 0 < area < area_range[0]:
            forward = 30   # too far, move the drone forward
        elif area > area_range[1]:
            forward = -30  # too close, move the drone backward
        else:
            forward = 0    # just good
        # upward/downward: whether center of the eyes is in upper or lower of the image
        if 0 < y < h/3:
            upward = 30    # too high, move the drone up
        elif y > h/2:
            upward = -30   # too low, move the drone down
        else:
            upward = 0     # just good
        # return
        self.prev_offset = x_offset
        return [0, forward, upward, yaw]

    def run(self):
        """Main program. Capture keyboard input in an infinite loop and
        translate to drone motion command or state mutation command."""
        self.prev_offset = 0
        speed = 75  # constant step size
        while not self.terminate:
            # default
            right = forward = upward = yaw = 0
            # start and stop the drone
            if getkey("p"):     # emergency shutdown
                self.drone.emergency()
            elif getkey("q"):   # land the drone
                self.drone.land()
            elif getkey("e"):   # take off
                self.drone.takeoff()
            # drone movement commands: multiple keys can be pressed at the same time
            if getkey("RIGHT"): right   += speed
            if getkey("LEFT"):  right   -= speed
            if getkey("UP"):    forward += speed
            if getkey("DOWN"):  forward -= speed
            if getkey("w"):     upward  += speed
            if getkey("s"):     upward  -= speed
            if getkey("d"):     yaw     += speed  # yaw to right
            if getkey("a"):     yaw     -= speed  # yaw to left
            # toggle commands
            if getkey("g"):   # toggle gesture controlled drone
                self.gestures = not self.gestures
                self.followme = self.findfaces = False
            if getkey("k"):   # toggle eye-tracking flying
                self.followme = not self.followme
                self.gestures = self.findfaces = False
            if getkey("l"):   # toggle counting faces on camera
                self.findfaces = not self.findfaces
                self.gestures = self.followme = False
            if getkey("f"):   # toggle display of camera window
                self.streaming = not self.streaming
                if self.streaming:
                    self.drone.streamon()
                else:
                    self.drone.streamoff()
                    cv2.destroyAllWindows()
            if getkey("r"):   # toggle recording video to file
                if self.recording:
                    self.recording = self.streaming = False
                    self.end_recording()
                else:
                    self.recording = self.streaming = True
                    self.start_recording()
            # if we need the camera, make sure it turned on
            if self.gestures or self.followme or self.findfaces:
                self.drone.streamon()
                self.recording = self.streaming = True
            # overwrite movement if motion can be controlled by camera
            if self.streaming:
                cv_movement = self.refresh_camera()
                if cv_movement is not None:
                    right, forward, upward, yaw = cv_movement
            # adjust the drone, and wait for 50ms for next 
            self.drone.send_rc_control(right, forward, upward, yaw)
            time.sleep(0.05)

if __name__ == "__main__":
    DroneControl().run()
\end{minted}

\nocite{*}
\printbibliography

@online{opencv,
    url = {https://github.com/opencv/opencv/tree/master/data/haarcascades},
    title = {Haar Cascade Classifiers},
    author = {OpenCV},
    year = 2020,
}

@online{hand,
    url = {https://github.com/kinivi/hand-gesture-recognition-mediapipe},
    title = {Hand Gesture Recognition using MediaPipe},
    author = {Nikita Kiselov},
    addendum = {This is a sample program that recognizes hand signs and finger gestures with a simple MLP using the detected key points. Handpose is estimated using MediaPipe},
    year = 2021,
}

@online{ytdji,
    url = {https://www.youtube.com/watch?v=URSCmWMIkE4},
    note = {YouTube video},
    title = {{DJI} Tello Face Tracking},
    author = {Shreyas Sharma},
    month = May,
    year = 2021,
}

@online{tello,
    url = {https://www.ryzerobotics.com/tello/downloads},
    year = 2022,
    author = {{Ryze Tech}},
    title = {Tello downloads},
}

@online{face,
    url = {https://github.com/murtazahassan/Drone-Face-Tracking/},
    %https://github.com/murtazahassan/Drone-Face-Tracking/blob/master/utlis.py
    title = {Drone Face Tracking},
    author = {Murtaza Hassan},
    year = 2020,
}

@inproceedings{haar,
    title = {Rapid Object Detection using a Boosted Cascade of Simple Features},
    author = {Paul Viola and Michael Jones},
    booktitle = {Proc. CVPR},
    year = 2001,
}

@article{haar2,
    author = {Paul Viola and Michael Jones},
    title = {Robust real-time face detection},
    journal = {International Journal of Computer Vision},
    volume = 57,
    number = 2,
    pages = {137--154},
    year = 2004,
}

\end{document}